\documentclass[11pt,twoside]{article}
\usepackage{./asp2014}

\aspSuppressVolSlug
\resetcounters

\begin{document}

\title{Effective Temperatures of Selected Main-sequence Stars with Most Accurate Parameters}
\author{F. Soydugan$^{1}$, Z. Eker$^{2}$, E. Soydugan$^{1}$, S. Bilir$^{3}$,  E.Yaz G\"{o}k\c{c}e$^{3}$, I. Steer$^{4}$, M. T\"uys\"uz$^{1}$, T. \c{S}eny\"uz$^{1}$ and O. Demircan$^{5}$
\affil{$^1$Department of Physics, Faculty of Arts and Sciences, \c{C}anakkale Onsekiz Mart University, \c{C}anakkale, Turkey; \email{fsoydugan, esoydugan@comu.edu.tr}}
\affil{$^2$Department of Space Sciences and Technologies, Faculty of
Sciences, Akdeniz University, Antalya, Turkey;
\email{eker@akdeniz.edu.tr}}
\affil{$^3$Department of Astronomy and Space Sciences, Faculty of Science, Istanbul University, Istanbul, Turkey; \email{sbilir, esmayaz@istanbul.edu.tr}}
\affil{$^4$NASA/IPAC Extragalactic Database, Pasadena, California, USA, Turkey; \email{ian@colosseum.com}}
\affil{$^5$Department of Space Sciences and Technologies, Faculty of Arts and Sciences,
\c{C}anakkale Onsekiz Mart University, Turkey; \email{demircan@comu.edu.tr}}
}

\paperauthor{Sample~Author1}{Author1Email@email.edu}{ORCID_Or_Blank}{Author1 Institution}{Author1 Department}{City}{State/Province}{Postal Code}{Country}
\paperauthor{Sample~Author2}{Author2Email@email.edu}{ORCID_Or_Blank}{Author2 Institution}{Author2 Department}{City}{State/Province}{Postal Code}{Country}
\paperauthor{Sample~Author3}{Author3Email@email.edu}{ORCID_Or_Blank}{Author3 Institution}{Author3 Department}{City}{State/Province}{Postal Code}{Country}

\begin{abstract}
In this study, the distributions of the double-lined detached binaries (DBs) on the planes of mass-luminosity, mass radius and mass-effective temperature have been studied. We improved the classical mass-luminosity relation based on the database of DBs by Eker et al. (2004a). With accurate observational data available to us, a method for improving effective temperatures for eclipsing binaries with accurate masses and radii were suggested.
\end{abstract}

\section{Data and Calibrations}
Among the eclipsing binary stars, detached double-lined spectroscopic binaries (DBs) are the most important sources in order to obtain the main physical properties of stars precisely. They are also valuable to test the models of stellar structure and evolution and also improve basic relationships between absolute parameters of stars. For these reasons, DBs were preferred for this study to present the distributions of main parameters, improve mass-luminosity relation (MLR) and also suggest an alternative method to determine effective temperatures of eclipsing binaries.

A calibration sample was compiled by selecting main-sequence stars with accurate masses, radii and effective temperatures from the recent database of DBs by Eker et al. (2014a). We selected 298 stars with masses and radii accurate to $\leq$ 3\% among 257 DBs in the catalogue. Using the data, we improved the classical MLR for the full mass range (0.2 $<$ M/M$_{\odot}$ $\leq$ 31) and also four mass domains, as indicated in Fig. 1. The mass-radius (MR) and mass-effective temperature relations are also investigated. Due to the evolutionary effects (M $>$1M$_{\odot}$), a single function to represent MR relation would be meaningless. The evolution on the main-sequence band is not seen obvious on the M - T$_{eff}$ diagram. Therefore, it was thought that the effective temperature could be re-calculated by using the updated MLR in different stellar mass domains.

\section{Conclusions}

We improved the classical MLR for different mass domains with linear equations. Our investigation of $\alpha$ on classical mass-luminosity relation (L$\propto$M$^{\alpha}$) for various mass ranges on the main-sequence was presented. A practical method was suggested to compute effective temperature if mass and radius of a star are known. With the calculated temperatures, the stellar temperature evolution within the main-sequence band for stars with M > 1M$_{\odot}$ is clearly visible on the $M-T_{eff}$ diagram. The details of the study and also the updated classical mass-luminosity relations are given by Eker et al. (2014b).

\articlefigure[width=.7\textwidth]{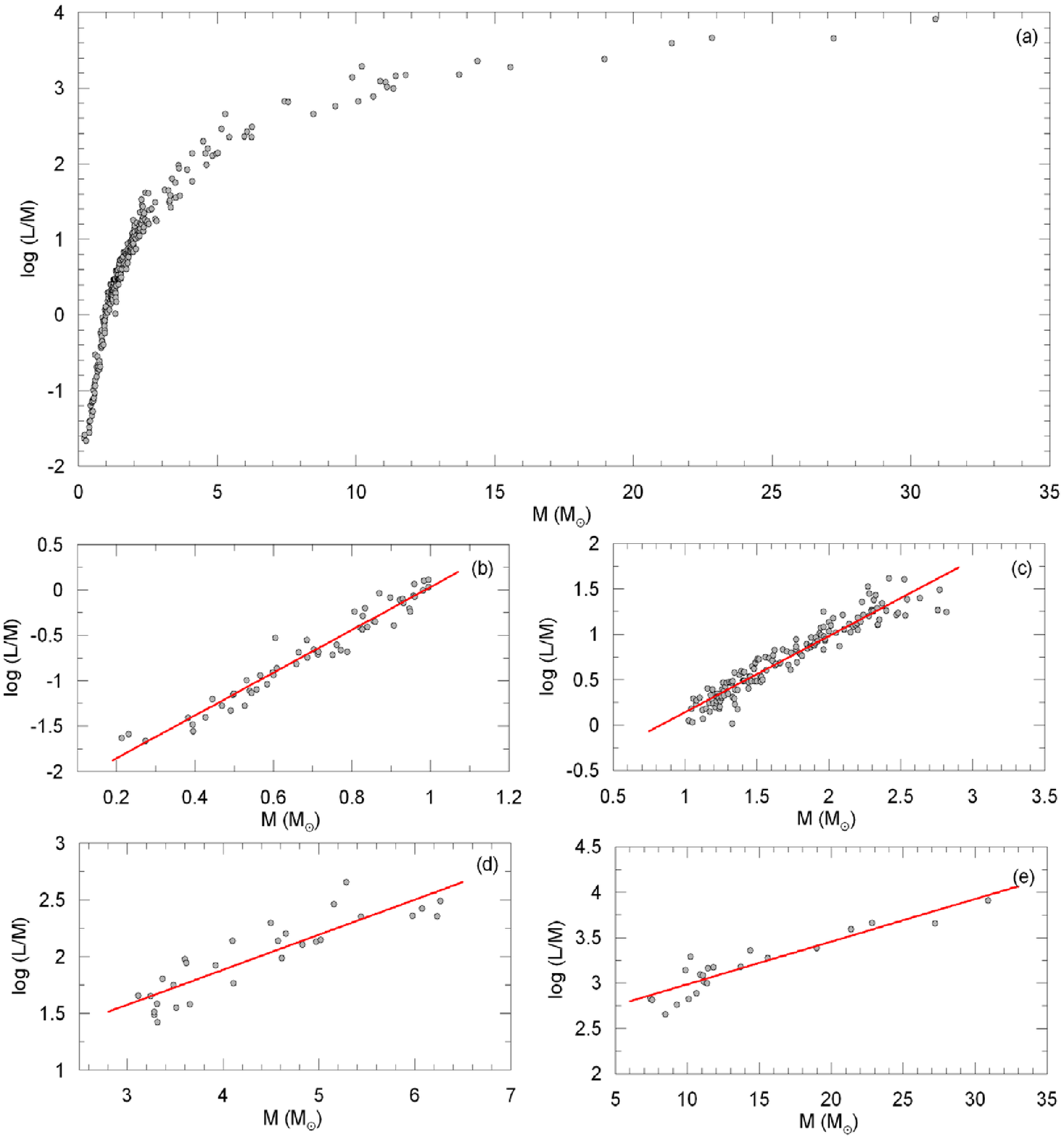}{}{The M-L diagram for the whole sample and also four different mass intervals.}

\acknowledgements This work has been supported in part by the Scientific and Technological Research Council (T\"{U}B\.{I}TAK) grant numbers 106T688 and 111T224.

\end{document}